# Modelling the effect of fiber distribution on the transverse mechanical characteristics of unidirectionally reinforced continuous-fiber composite

Sergejs Tarasovs[a], Janis Modniks[a], Andrea Bercini Martins[b], Christina Scheffler[bc], Janis Andersons[a]

[a] Institute for Mechanics of Materials, Faculty of Science and Technology, University of Latvia, Jelgavas St. 3, LV-1004, Riga

[b] Department of Fiber Engineering, Leibniz Institute of Polymer Research Dresden e.V., 01069 Dresden, Germany

[c] Institute of Construction Materials, Dresden University of Technology, 01062 Dresden, Germany

Corresponding author: sergejs.tarasovs@lu.lv (Sergejs Tarasovs)

**Abstract**

This study investigates the influence of fiber spatial distribution on the transverse mechanical properties of unidirectionally reinforced continuous-fiber composites. A Swelling & Random Migration algorithm was employed to generate representative volume elements with controlled fiber arrangements, ranging from clustered to equilibrium configurations. Finite element homogenization with periodic boundary conditions was used to estimate effective elastic properties. To characterize fiber randomness and assess statistical equivalence with experimental microstructures, several descriptors are employed, including nearest neighbor distance, Ripley's K-function, pair distribution function, and local fiber volume fraction. Results reveal that, at constant fiber volume fraction, clustered fiber distributions yield significantly higher transverse stiffness but lower transverse tensile strength compared to the equilibrium distributions. For glass/epoxy composites, transverse stiffness varies by up to 20% depending on the degree of fiber clustering. A single scalar descriptor, the mean nearest neighbor distance, was shown to efficiently characterize sufficiently random fiber distributions: effective stiffness decreases, whereas transverse tensile strength increases linearly with mean nearest neighbor distance. The findings highlight the critical role of microstructural characteristics in tailoring composite performance and provide a robust framework for predictive modeling of fiber reinforced materials.

**Keywords**: unidirectional fiber composites, effective properties, representative volume element, microstructure, statistical descriptors

# 1. Introduction

Continuous-fiber unidirectionally reinforced (UD) composites, as the designation implies, possess aligned reinforcing fibers. The spatial distribution of fibers, however, may differ among UD composites, both due to their production technology [1], which imparts systematic variation in fiber location, and stochastic effects leading to random variability in local fiber volume fraction and inter-fiber spacings. A UD composite morphology with discernible fiber bundles, having a higher volume fraction than the mean of the composite and separated by regions of matrix, is typical for composites produced by filament winding and filament printing [2]. Although UD composites manufactured from prepregs tend to possess an apparently uniform random distribution of fibers [3, 4], resin-rich zones between adjacent prepreg plies may be present. Fiber clustering and matrix pockets are commonly observed in composites regardless of their production method. Even in composite areas with relatively uniform fiber content, their positions are stochastic [5-8].

Fiber surface treatment can also affect the spatial distribution of fibers within composite materials, with the imposition of limitations on how close neighboring fibers can be placed. While the sizing layer on fibers is usually relatively thin, measuring in tens to hundreds of nanometers [9, 10], plasma coating thicknesses of up to 10 microns have been considered [11]. Such coatings would act as fiber spacers in a UD composite.

For the quantitative characterization of disorder in the fiber arrangement in a UD composite, several statistical descriptors have been considered. Most frequently used descriptors include Ripley's $K$-function, as well as nearest neighbor and pair distributions [1, 12]. Nearest neighbor distance and orientation distributions characterize the proximity of the closest fiber to the given one, and its relative angular position in a selected coordinate system, respectively. Ripley's $K$-function yields information on fiber distribution at larger distances; the function $K(r)$ is defined as the number of fibers located within a distance $r$ from a selected fiber normalized by the area density of fibers (i.e., average number of fibers per unit area). Pair distribution function $g(r)$ gives the probability of finding a fiber within a hoop with a center coinciding with another fiber, internal radius $r$, and a specified thickness; it is proportional to the derivative of Ripley's $K$-function $g(r) \sim dK(r)/dr$. Fiber distribution can also be described via such characteristics of Voronoi polygons, formed by Dirichlet tessellation associated with fiber centers, as distributions of polygon area (or local fiber volume fraction), number of sides, etc. [1, 13-15]. Voronoi tessellation can also be used to define natural neighbors of a given fiber and calculate the Voronoi neighbors distance distribution.

Statistical descriptors enable quantitative characterization of fiber distribution and evaluation of agreement of simulated fiber distributions with the experimental ones [1, 3, 4, 12, 13]. For a simulated UD composite microstructure with random fiber distribution, the statistical descriptors of which closely agree with those derived from micrographs of the actual UD composites, the elasticity characteristics obtained by numerical homogenization are also shown to be in good agreement with the measured ones [3, 4]. The modelling accuracy in this case is commensurate with that achieved by using fiber distribution derived directly from composite micrographs in elastic property prediction [2].

Fiber distribution strongly affects stress concentration in the matrix and fiber/matrix interface, thus governing damage initiation and strength of UD composites, see, e.g., [1, 7, 16, 17].

Non-uniform fiber placement also affects the elastic properties of UD composites. While the longitudinal Young's modulus of a UD composite is virtually unaffected by fiber distribution [3, 18], numerical simulations suggest that the transverse modulus for random fiber distribution is bounded by the moduli of periodic square and hexagonal fiber arrays [3, 18-20]. Transverse shear modulus for random UD fiber placement is found to either slightly exceed those of square and hexagonal fiber arrangement [3, 18, 20] or to be bounded by them [19], apparently depending on the distribution characteristics.

The aim of this study is to elucidate the relation between fiber distribution, characterized by the relevant descriptors, and the elastic properties of the UD composite. For that, fiber distributions are generated using Swelling & Random Migration (SRM) algorithm [21], varying systematically the minimum fiber spacing and degree of randomness. The elasticity parameters of the respective UD composites are estimated using a finite element homogenization framework with periodic representative volume elements (RVE) and periodic boundary conditions.

# 2. Representative Volume Element

## 2.1. Generation of random fiber distributions

A statistically representative volume element (RVE) containing randomly placed mono- and polydisperse circular fibers was generated using an in-house developed software implementing the Swelling & Random Migration (SRM) algorithm. This collective-rearrangement algorithm is capable of producing a wide range of microstructures with controlled degrees of randomness, from equilibrium-like configurations (maximally random, hard-core distributions) to strongly clustered arrangements characterized by compact fiber agglomerates and extended matrix-rich regions.

The SRM algorithm begins with $N$ non-overlapping fibers placed randomly within a unit square. Their radii are then increased gradually, and at each swelling step overlaps are resolved through cooperative random motion of the particles. This iterative process continues until the target fiber volume fraction is reached. The swelling rate and the magnitude of random migration serve as the primary control parameters governing the degree of clustering in the resulting microstructure. Periodicity of the RVE is enforced throughout the procedure by introducing periodic image particles on opposite boundaries of the domain. A detailed description of the SRM algorithm and its implementation can be found in [21].

Once the desired fiber volume fraction is achieved, the RVE window is translated randomly within the infinite periodic tiling of the particle arrangement. For each candidate position, the minimum distance between any fiber and the RVE boundary is evaluated. The window position that maximizes this minimum distance is selected as the final RVE configuration. This post-processing step prevents excessively small gaps between fibers and the RVE boundary, which would otherwise lead to severe element distortion during finite element meshing of the microstructure [22].

By varying the parameters of the RVE-generation algorithm, specifically, the admissible inter-particle distance and the intensity of random migration, the SRM procedure produces a wide spectrum of fiber arrangements. In the limiting cases, three distinctly different configurations can be obtained: equilibrium, clustered, and well-separated distributions.

A high random-migration intensity combined with a low swelling rate yields a highly disordered microstructure approaching an equilibrium (maximally random) distribution of fibers. Conversely, when the random-migration intensity is low, particles tend to remain in contact during swelling, forming chains or compact agglomerates reminiscent of the sticky-disk model [23]. Increasing the admissible inter-particle distance produces the opposite effect: fibers are prevented from approaching one another, resulting in well-separated configurations. At sufficiently large admissible distances, whose exact value depends on the target fiber volume fraction, the system undergoes local crystallization, forming regions with nearly hexagonal packing.

By gradually adjusting the admissible spacing and migration intensity, a set of intermediate microstructures can be generated. Fig. 1 illustrates representative equilibrium, clustered, and well-separated arrangements for a system of 200 fibers at a volume fraction of 0.65.

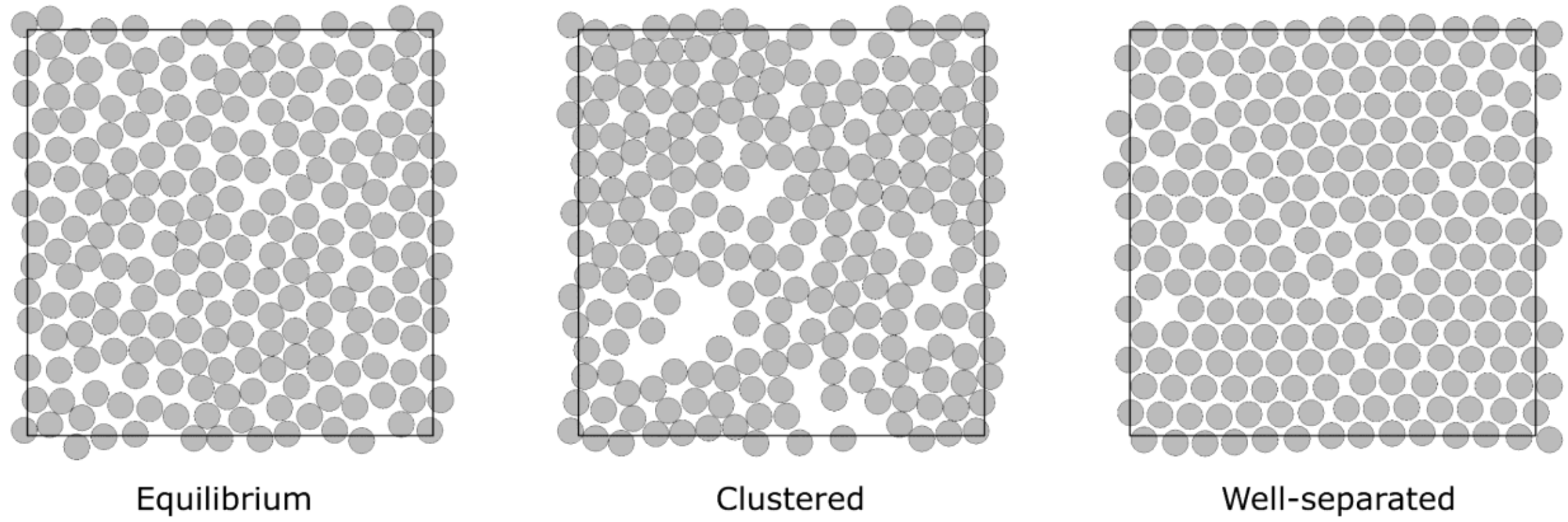

Fig. 1. Examples of equilibrium, clustered, and well-separated microstructures generated by the presented procedure for the volume fraction of fibers equal to 0.65.

An example of clustered configurations generated for 200 fibers at different volume fractions is shown in Fig. 2. At moderate fiber contents (30–50%), short chains of fibers and small compact clusters begin to form. As the volume fraction increases, these clusters progressively merge, eventually forming a single percolating agglomerate that spans the entire RVE. Even at a high fiber volume fraction of 70%, relatively large resin pockets remain present within the clustered configuration, frequently observed in real unidirectional composites.

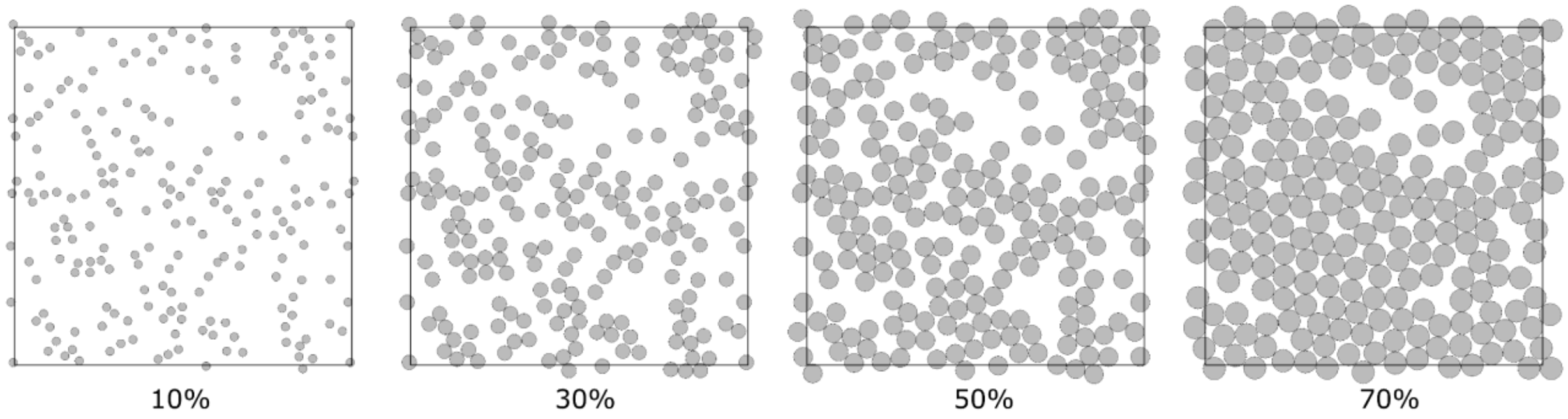

Fig. 2. Generation of a clustered configuration of fibers from a Poisson distribution of initial seed points.

## 2.2. Finite element mesh

The generated fiber arrangements were imported into Gmsh [24], where fibers intersecting the RVE boundary were clipped and a periodic finite element mesh was constructed. Six-node triangular elements with curved edges were employed to accurately represent the circular fiber geometry. For most simulations, the ratio of average element size in matrix regions (away from fiber boundaries) to fiber diameter was maintained between 0.1 and 0.2. A mesh-convergence study confirmed that an accuracy of 1% or better is achieved when this ratio does not exceed 0.3.

Local mesh refinement was applied in regions where fibers were nearly touching. In these zones, the element size was scaled proportionally to the minimum gap between adjacent fibers, as illustrated in Fig. 3. This refinement strategy prevents excessive element distortion and ensures accurate resolution of the displacement field in narrow matrix ligaments.

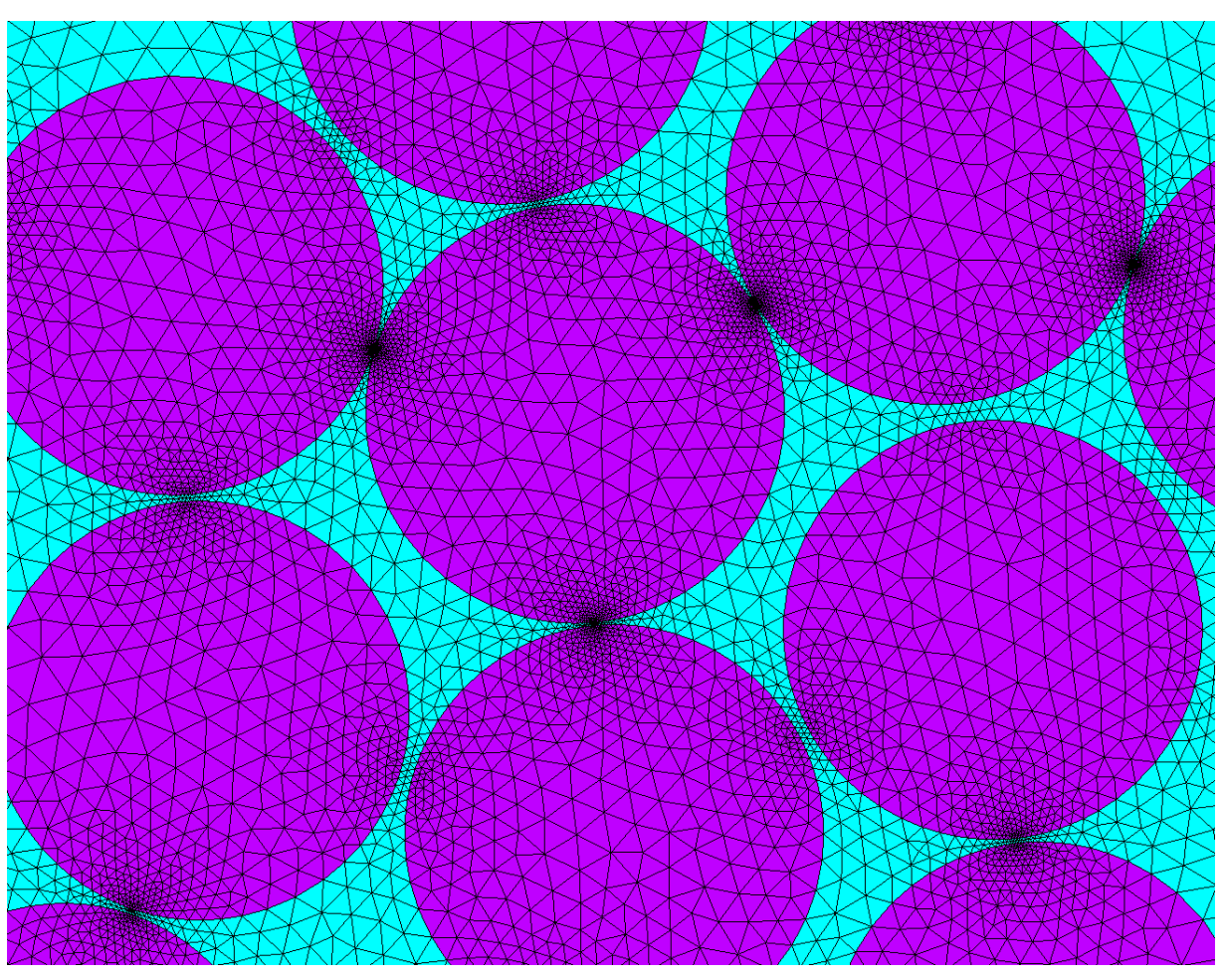

Fig. 3. Local mesh refinement between fibers in close contact

## 2.3. Size of Representative Volume Element

The choice of an appropriate RVE size is essential for balancing computational efficiency with accuracy. The RVE must be sufficiently large to capture the statistical features of the fiber distribution while remaining small enough to allow extensive parametric studies. For unidirectional fiber-reinforced composites, previous studies suggest that a characteristic RVE size satisfying $L/R \geq 40$, where $L$ is the RVE side length and $R$ is the fiber radius, is typically required for representativeness [25-27].

In [26] a micromechanical study of stress concentrations in composites was performed for several RVE sizes and fiber volume fractions of 40%, 50%, and 60%, using periodic boundary conditions. Multiple realizations were generated for each size, and the maximum von Mises stress in the matrix was recorded. It was found that the standard deviation of the peak matrix stress ceased to decrease for RVE size to fiber radius ratio $L/R \geq 40$, indicating that further enlargement of the RVE did not improve statistical stability.

A systematic investigation of RVE size requirements was performed in [27], where different criteria for statistical representativeness were examined. The authors concluded that $L/R \geq 40$ is necessary for reliable comparison of distance-based statistical descriptors, whereas $L/R \geq 30$ is sufficient to estimate effective elastic properties with an error below 10%.

In the present work, an RVE size of $L/R > 50$ was employed when comparing statistical descriptors of generated and experimental fiber distributions, ensuring robust characterization of spatial statistics. For the estimation of effective elastic properties across a range of generated microstructures and fiber volume fractions, RVEs with an average size ratio of approximately $L/R \approx 40$ were used.

# 3. Experimental

## 3.1. Materials

A commercial epoxy resin with an anhydride-based hardener was used at a weight ratio of 100:90, respectively. The used curing cycle involved heating to 80 °C for 4 hours, followed by post-curing at 145 °C for a further 4 hours. A commercially available E-Glass fiber from Owens Corning with a tex of 2400 and an average diameter of 24 μm was used as the reinforcing fiber.

## 3.2. Composite manufacturing

The UD composite panels with two different fiber volume fractions (target values of 50% and 65%) were manufactured using vacuum-assisted resin transfer moulding (VARTM). First, the dry fibers were wound around a metal core using a winding machine (IWT GmbH, model FW 322). The VARTM tool was developed at the Leibniz-Institut für Polymerforschung Dresden e. V., Germany. It is vacuum-tight and consists of two rigid mould halves and one metal core on which the wound preform is located (see Fig. 4). The tool was then closed and sealed with vacuum grease before being inserted into a laboratory hot press (Schwabenthan Polystat 300s). The tool was pressurized in the hot press to avoid resin leakage. The mold inlet was connected to a resin reservoir, while the outlet was connected to a resin trap and a vacuum pump. The resin and the hardener were mechanically mixed by a speed mixer (Hauschild Speed Mixer™ DAC 400 FV) at 1500/2000/2400/2250/1700 rpm for 0.12 min at each stage. For the infusion process, the mixture was degassed at ambient temperature and then heated at 40 °C before being forced into the cavity between the two mould halves to impregnate the fibers. The resin was pressurized to enhance its flow. The flow was stopped when the resin filled the resin trap. After the infusion, the composite panel was cured and post-cured at 80 °C and 145 °C for 4 hours, respectively.

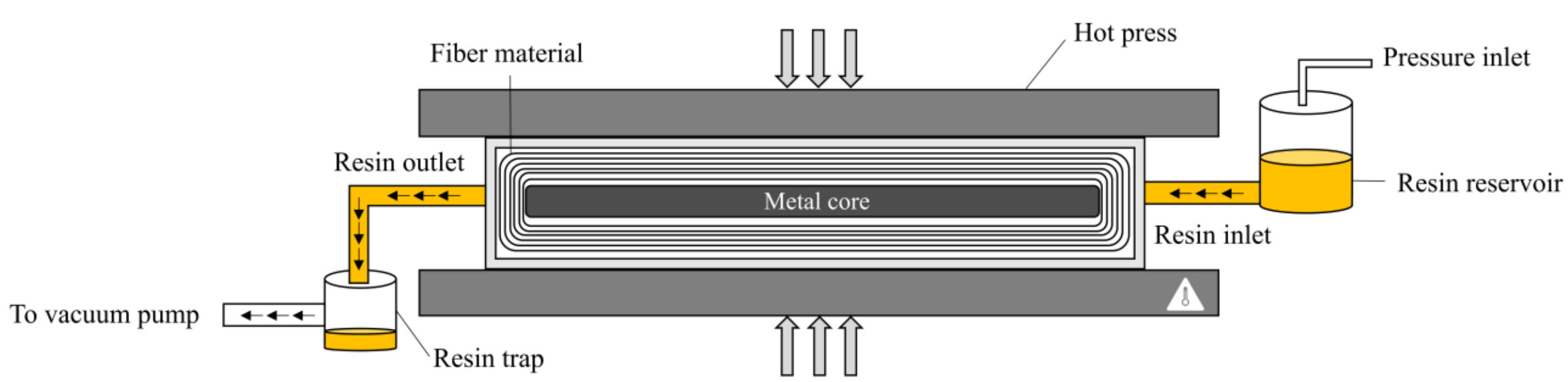

Fig. 4. Schematic drawing of the used Vacuum assisted resin transfer molding (VARTM)

### 3.3. Optical microscopy images

The composite panels were cut into small pieces using a cut-off saw machine (Mutronic® Diadisc 5200), and the cross-sections were exposed for polishing. Epoxy resin was cured around each composite piece to hold it in place during polishing. The polishing process began with P320 grinding paper and continued with P1200, followed by a 9 μm diamond suspension, a 3 μm diamond suspension, and a 0.06 μm alumina particle suspension (Buehler Ltd. Model Phoenix 4000). Optical microscopy images were taken using a Keyence VH-Z500 microscope.

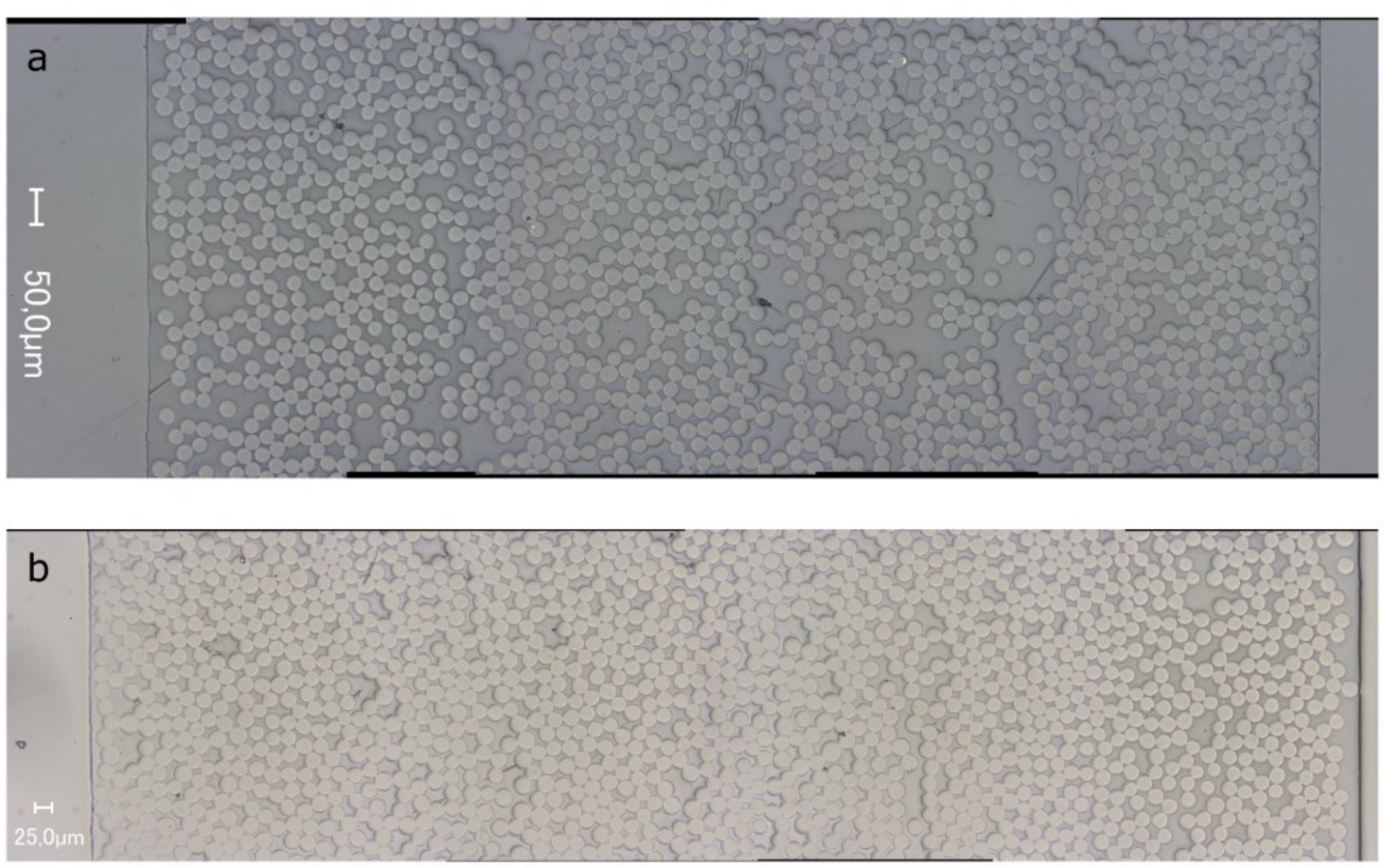

Fig. 5. Cross-sections of composite plates with fiber volume fractions of 54.7% (a) and 67.3 % (b).

The diameters and centroid positions of all individual fibers in the composite cross-sections shown in Fig. 5 were extracted using ImageJ. These measurements were subsequently used to reconstruct the corresponding fiber arrangements in CAD and finite element models. Fiber volume fractions for the analyzed cross-sections were determined from the reconstructed CAD geometries by cropping the outer regions of the models to eliminate edge effects and computing the total fiber area within the resulting observation window.

## 4. Statistical descriptors

A variety of statistical descriptors are used to quantify the spatial distribution of particles in composite materials. Among the most widely employed measures are the nearest neighbor distance distribution, Ripley's $K$-function, and the pair distribution function $g(r)$ [1, 12]. Additional descriptors, such as the distribution of Voronoi-based neighbors or the local fiber volume fraction, are also frequently utilized [15]. These metrics are computationally efficient, straightforward to interpret, and provide practical alternatives to higher-order $n$-point correlation functions. In this study, these statistical descriptors were used to assess the statistical equivalence between randomly generated fiber distributions in the RVEs and experimentally observed fiber arrangements in unidirectional composite materials.

### 4.1. Nearest neighbor distribution

The nearest neighbor distribution is defined as the probability density function (PDF) of the distances from each fiber to its closest neighboring fiber. This descriptor characterizes short-range spatial interactions within the microstructure. In particular, the shape of the PDF reflects the degree of local ordering or clustering: narrow distributions with small characteristic distances indicate tightly packed or clustered regions, whereas broader distributions with larger characteristic distances correspond to more dispersed or well-separated fiber arrangements.

### 4.2. Voronoi neighbors distribution

Using Voronoi tessellation, the natural neighbors of a fiber are defined as all fibers whose corresponding Voronoi cells share a common edge [25]. For each fiber, the distances to all of its Voronoi neighbors can be computed, and the probability density function (PDF) of these distances provides a descriptor that captures both short-range and long-range spatial characteristics of the microstructure. In particular, the presence of large matrix-rich regions, commonly observed in fiber-reinforced composites, manifests as a long tail in the Voronoi-neighbor distance PDF, reflecting the existence of fibers separated by large gaps.

### 4.3. Ripley's $K$-function

Ripley's $K$-function quantifies the spatial distribution of fibers by measuring the expected number of fiber centers located within a distance $h$ of an arbitrary fiber, normalized by the average areal fiber density. It is defined as

$$K(h) = \frac{A}{N^2} \sum_{i=1}^{N} n_i(h), \tag{1}$$

where $A$ is the observation area, $N$ is the total number of fibers within this area, and $n_i(h)$ is the number of fiber centers located inside a circle of radius $h$ centered at the "$i$"-th fiber. For periodic RVEs, edge effects are eliminated by taking the entire RVE as the observation window and computing $n_i(h)$ using periodic images of fibers surrounding the domain [28]. For experimental micrographs, where the fiber arrangement is non-periodic, a cropped region is used as the observation area, and $n_i(h)$ for each fiber is evaluated with respect to the full micrograph.

For a Poisson point process, representing an idealized random distribution, the analytical form of Ripley's $K$-function is given by [12]

$$K_P(h) = \pi h^2 \tag{2}$$

Deviations of measured $K(h)$ from this reference curve indicate non-random spatial structure: values above $K_{\mathrm{P}}(h)$ correspond to clustered fiber arrangements, whereas values below $K_{\mathrm{P}}(h)$ reflect more regular or dispersed configurations [1].

### 4.4. Pair distribution function

The pair distribution function $g(h)$ describes the probability of finding the center of a fiber within a hoop of inner radius $h$ and thickness $\Delta h$. It is defined as

$$g(h) = \frac{1}{2\pi h\rho \cdot \Delta h} \frac{1}{N} \sum_{i=1}^{N} n_i(h), \tag{3}$$

where $\rho = N/A$ is the density of fibers in observation area $A$. The pair distribution function is directly related to Ripley's $K$-function through

$$g(h) = \frac{1}{2\pi h} \frac{dK(h)}{dh}. \tag{4}$$

For periodic fiber arrangements, $g(h)$ exhibits distinct peaks at characteristic inter-fiber distances corresponding to the underlying lattice structure. As $h$ increases, the function $g(h)$ approaches 1, reflecting the loss of spatial correlation at large separations. The same considerations as for $K(h)$ can be used for handling edge effects in periodic and non-periodic RVEs.

### 4.5. Local volume fraction

The local fiber volume fraction (LVF), or local area fraction, is defined as the ratio of the fiber cross-sectional area to the area of its corresponding Voronoi cell. This descriptor provides a localized measure of how densely fibers are packed within the microstructure. Previous studies have shown that the LVF distribution is highly informative for characterizing spatial heterogeneity in composite cross-sections. In particular, the standard deviation of the LVF distribution is sensitive to clustering effects [29], with broader distributions indicating stronger local variations in fiber packing and the presence of matrix-rich regions.

## 5. Effective elastic properties

For the estimation of the effective elastic properties, a two-dimensional generalized plane strain finite element model with periodic boundary conditions [30] was employed. This formulation enables the evaluation of all in-plane and out-of-plane effective elastic constants of the unidirectional composite, with the exception of the axial shear moduli. To fully characterize the homogenized stiffness response, four linearly independent macroscopic loading cases were applied to the RVE: in-plane uniaxial tension in the two orthogonal transverse directions, in-plane shear, and axial (out-of-plane) tension.

In the case of a periodic RVE microscopic displacement field has to fulfill periodicity conditions [31]

$$u(\boldsymbol{x} + \boldsymbol{L}) = u(\boldsymbol{x}) + \langle \boldsymbol{\epsilon} \rangle \cdot \boldsymbol{L}, \tag{5}$$

where $\boldsymbol{x}$ is a radius vector of a node on a given face, $\boldsymbol{L}$ is a translation vector that connects a pair of constrained nodes, and $\langle \boldsymbol{\epsilon} \rangle$ is a volume average of the strain field. Similarly, continuity of normal stress and shear components of stress field on opposite boundaries of an RVE leads to a traction anti-periodic condition:

$$t(\boldsymbol{x}+\boldsymbol{L}) = -t(\boldsymbol{x}). \quad (6)$$

If a variational framework is used for homogenization, the application of kinematic periodic boundary conditions guarantees the uniqueness of the solution and automatic fulfillment of anti-periodic traction boundary conditions [31, 32].

For a rectangular RVE aligned with the global coordinate axes, the averaged engineering stress and strain components can be obtained directly from the finite element solution. Macroscopic strains and stresses can be computed by dividing displacements or reaction forces at the master node by the corresponding RVE side length/area. These averaged quantities enable the evaluation of the components of the two-dimensional compliance matrix $s_{ij}$ (in Voigt notation) as

$$s_{ij} = \frac{\varepsilon_i}{\sigma_j}; \quad i,j = 1,2,6, \quad (7)$$

where $\sigma_j$ are the average stress components and $\varepsilon_i$ are the average global strain components.
The compliance matrix is related to Young's moduli $E_i$, shear moduli $G_{ij}$, and Poisson's ratios $\nu_{ij}$ via

$$s_{ij} = \begin{bmatrix} 1/E_x & -\nu_{yx}/E_y & 0 \\ -\nu_{xy}/E_x & 1/E_y & 0 \\ 0 & 0 & 1/G_{xy} \end{bmatrix}. \quad (8)$$

For a generalized plane strain finite element formulation, the axial stiffness and out-of-plane Poisson's ratios can be evaluated in the same manner as the in-plane properties. However, the effective axial shear moduli cannot be obtained from a two-dimensional model and require a full three-dimensional finite element analysis. The commercial finite element code ANSYS was used to compute the effective elastic properties, employing a custom APDL script to impose periodic boundary conditions by pairing opposite nodes on the RVE boundaries.

For reconstructed RVEs with non-periodic fiber arrangements, mixed boundary conditions were applied [33]. In this scheme, displacement boundary conditions are prescribed along the loading direction, while the remaining RVE boundaries are left traction-free. The resulting apparent elastic moduli lie between the bounds obtained by displacement and traction boundary conditions, and typically provide accurate estimates of the effective properties for non-periodic RVEs [34].

# 6. Results and discussions

## 6.1. Generation of statistically equivalent microstructures

A key requirement for any algorithm intended to generate random fiber distributions is its ability to reproduce microstructures that are statistically equivalent to those observed in real composite materials. In this work, the SRM algorithm was used to create RVEs with a wide range of fiber arrangements by systematically varying the swelling rate and the intensity of random migration.

The statistical descriptors of the generated microstructures were then compared with those extracted from the experimental images (Fig. 5). The results demonstrate that the algorithm successfully reproduces the essential spatial features of real fiber distributions. Quantitative comparisons of the nearest neighbor, Voronoi-neighbor, Ripley's $K$-function, and pair-distribution descriptors for simulated and experimental microstructures are presented in Fig. 6 and Fig. 7. The corresponding local fiber volume fraction distributions for both simulated RVEs and experimental micrographs are shown in Fig. 8.

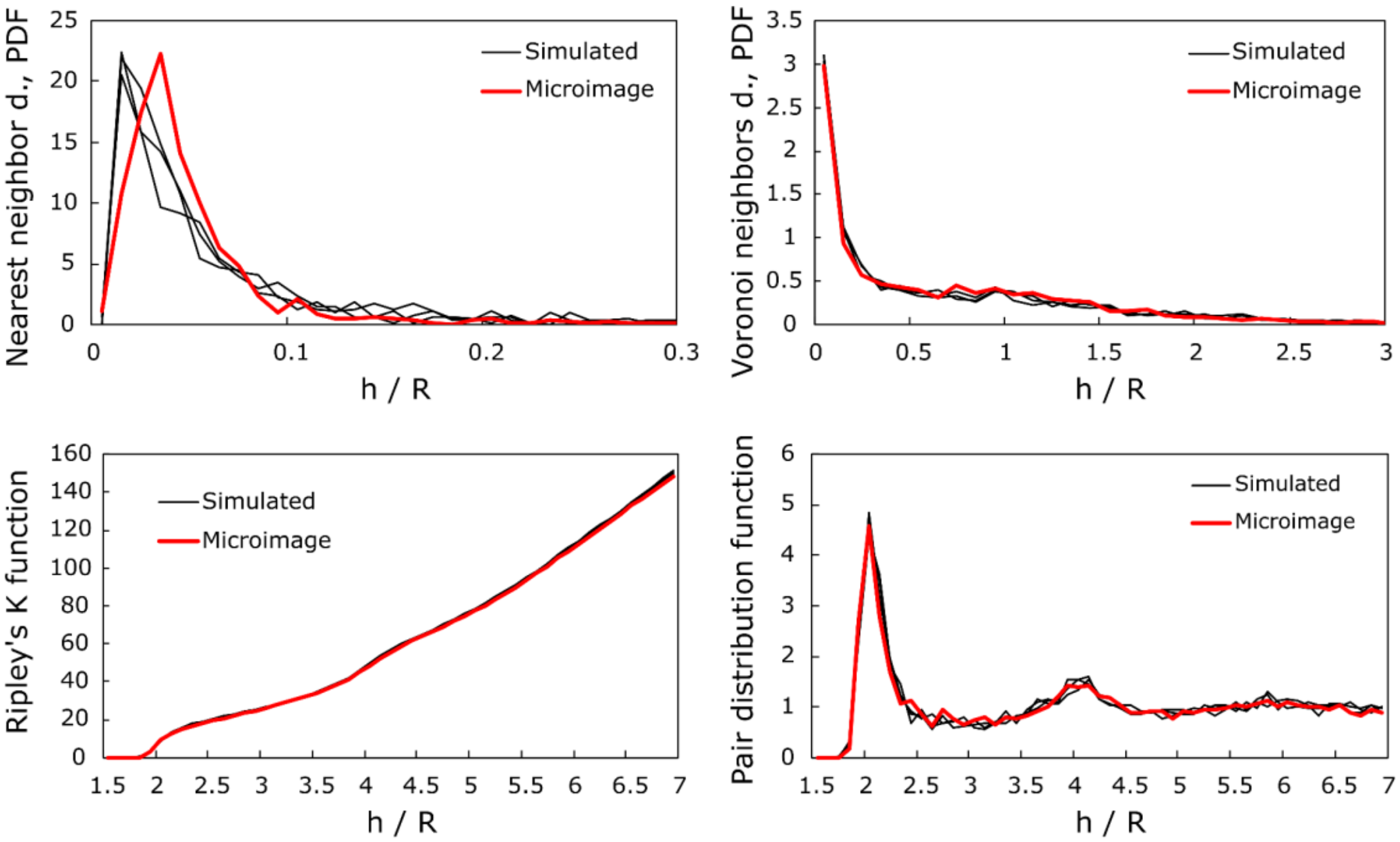

Fig. 6. Simulated and experimental statistical descriptors for a composite sample with a fiber volume fraction equal to 54.7%.

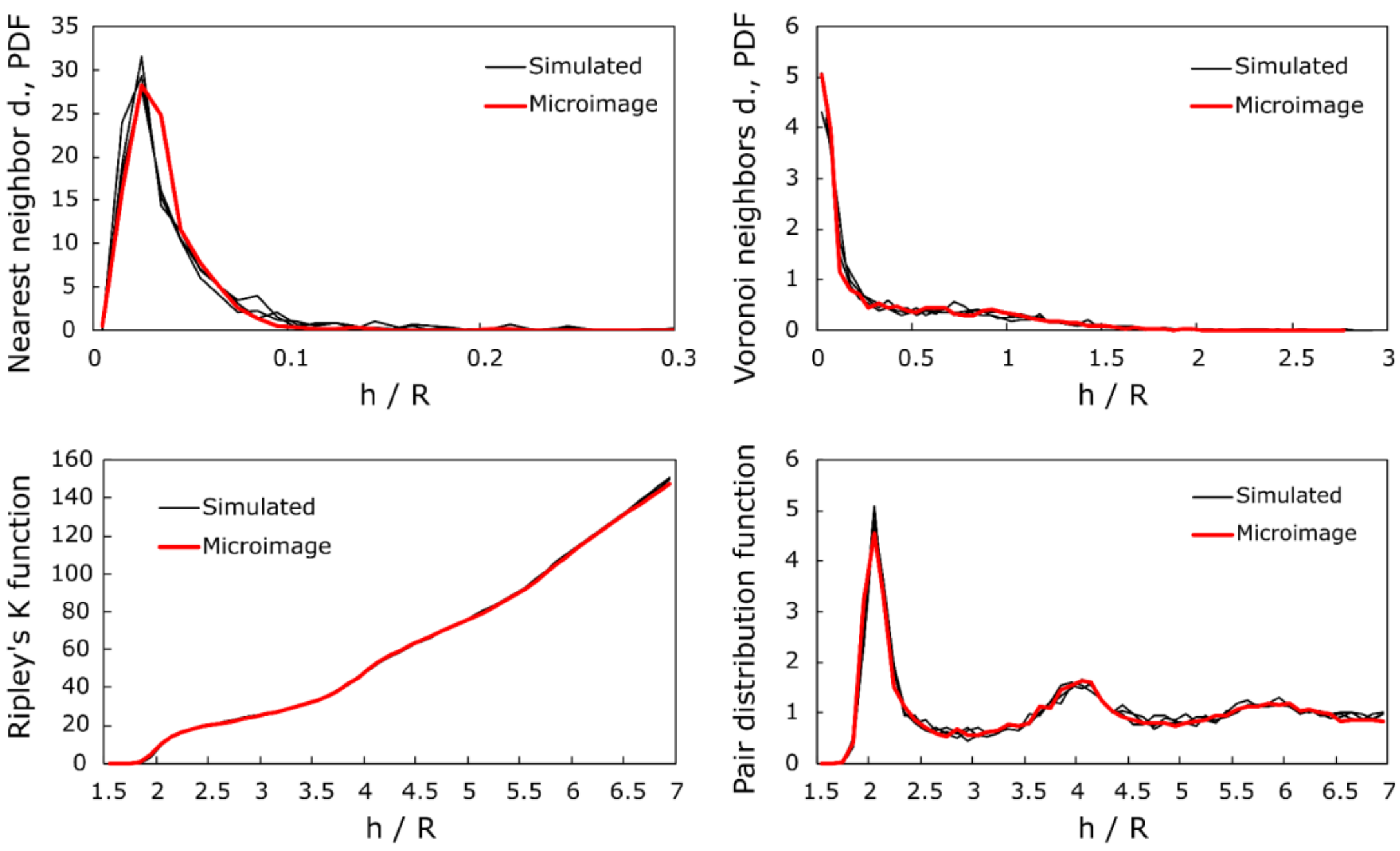

Fig. 7. Simulated and experimental statistical descriptors for a composite sample with a fiber volume fraction equal to 67.3%.

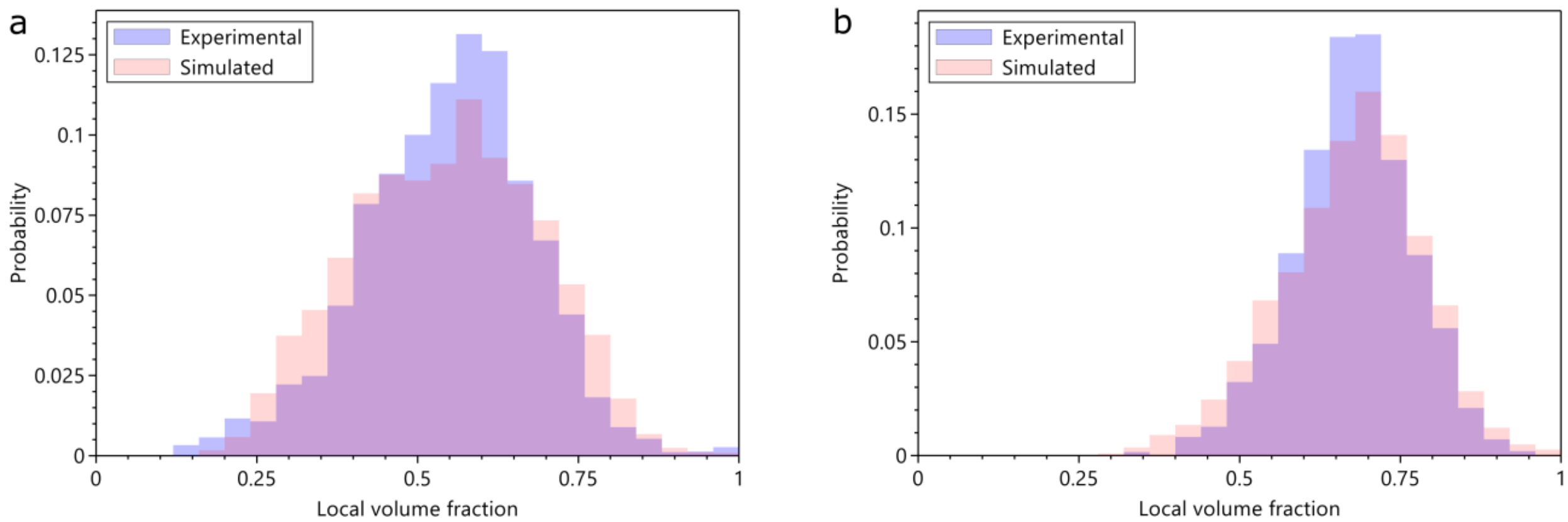

Fig. 8. Simulated and experimental local fiber volume fraction distribution for a composite sample with fiber volume fractions equal to 54.7% (a) and 67.3% (b).

For optimal statistical equivalence, the generated microstructures should incorporate polydisperse fibers whose diameter distribution matches that of the experimental material. In this work, fiber diameters were measured from the micrographs shown in Fig. 5 using ImageJ, with approximately 1000–1500 fibers analyzed for each sample. The measured diameter distributions were approximated by a log-normal distribution, and the corresponding distribution parameters for each micrograph were used to initialize the fiber sizes in the numerical generation algorithm.

For polydisperse fibers, the surface-to-surface distance between fibers was used when computing the nearest neighbor and Voronoi neighbor distributions, rather than the more common center-to-center distance [13]. This approach provides a more accurate characterization of local spacing in the presence of diameter variability. The average fiber radius was used to normalize the statistical descriptors presented in Fig. 6 and Fig. 7.

A slight discrepancy was observed in the nearest neighbor distribution at very small separations. This deviation is attributed to the limited resolution of the micrographs in Fig. 5, which affects the accuracy of distance measurements for nearly touching fibers. Examples of statistically equivalent generated microstructures containing approximately 470 and 620 fibers, corresponding to fiber volume fractions of 54.7% and 67.3%, respectively, are shown in Fig. 9.

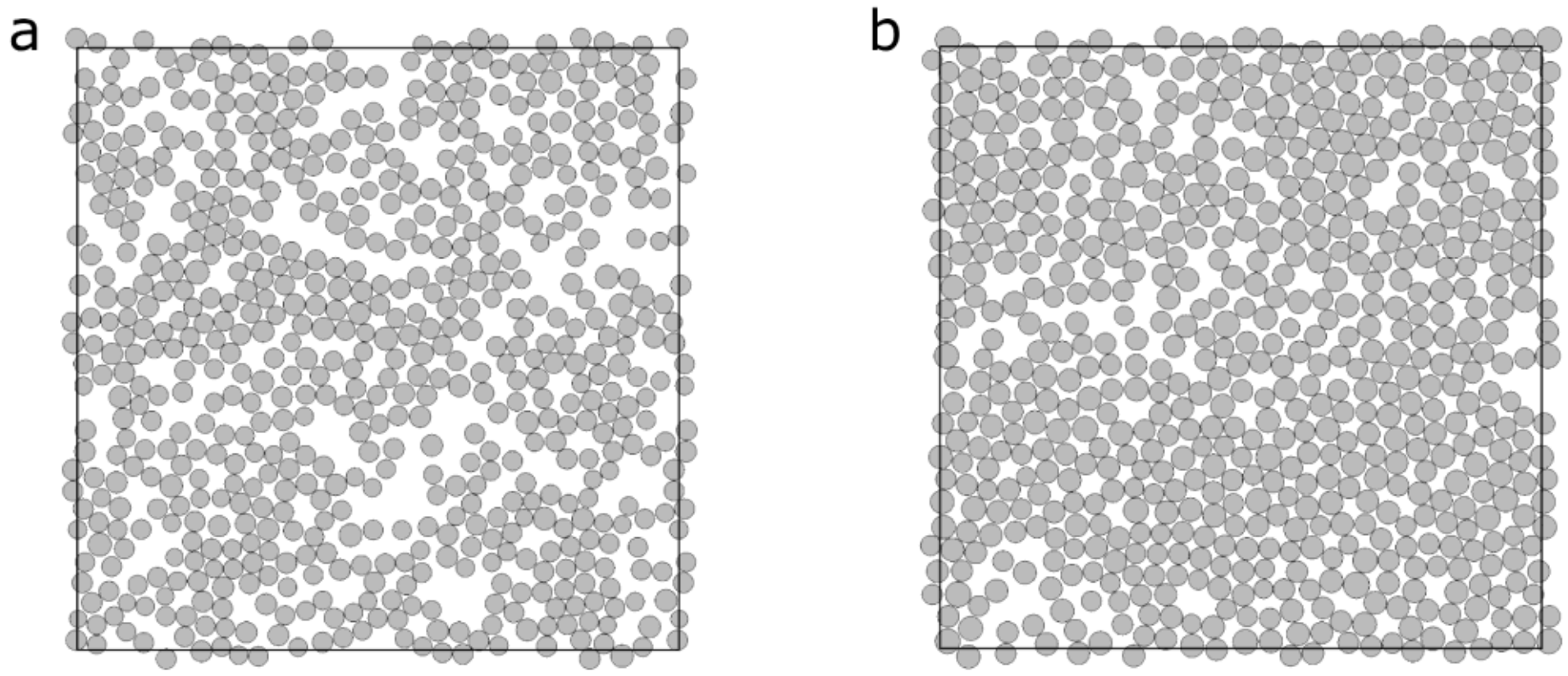

Fig. 9. Examples of generated RVEs with volume fractions of 54.7% (a) and 67.3% (b).

To further assess the similarity between real and generated microstructures, the effective elastic properties were computed for both reconstructed and simulated RVEs using identical constituent material properties for the fibers and matrix. In these analyses, the Young's modulus of the matrix and fibers was taken as 3.24 GPa and 82 GPa, respectively, with corresponding Poisson's ratios of 0.34 and 0.22.

For each fiber volume fraction, three independent random distributions were generated using the pre-determined SRM parameters. The Young's moduli were evaluated in two orthogonal transverse directions for every RVE, and the averaged values are reported in Table 1. For the reconstructed microstructures, the mean stiffness obtained from the two loading directions is presented. The results confirm that the SRM algorithm is capable of generating fiber arrangements that are statistically equivalent to those observed in real composites, and yield effective elastic properties in close agreement with those of the reconstructed microstructures.

Table 1 Comparison of effective stiffness for reconstructed and simulated distributions of fibers

| **Fiber volume fraction** | **Effective stiffness, GPa** | |
|---|---|---|
| | Reconstructed | Simulated |
| 54.7% | 11.94 ± 0.11 | 11.85 ± 0.11 |
| 67.3% | 17.71 ± 0.36 | 17.43 ± 0.18 |

## 6.2. Influence of interparticle distance

One of the important parameters in generating random fiber distributions is the minimum allowable surface-to-surface distance between fibers. Although real composites may contain fibers that are nearly touching, such configurations introduce numerical challenges in finite element simulations. Very small inter-fiber gaps require substantial mesh refinement, increase computational cost, and may lead to distorted or degenerate elements. For this reason, a minimum admissible spacing is typically imposed in RVEs used in numerical simulations.

To examine how this parameter influences effective elastic properties, two series of RVEs were generated in which the minimum fiber spacing was gradually increased. The spacing ranged from a small value of $0.0001R$ up to the maximum feasible distance for the specified fiber volume fraction, beyond which the microstructure transitions into a quasi-regular configuration approaching hexagonal packing.

In the first series, a minimal migration rate was used, producing microstructures with pronounced clustering. In the second series, a significantly higher migration rate was applied, resulting in equilibrium-like fiber distributions. Together, these two sets of RVEs represent smooth transitions from clustered or equilibrium arrangements toward well-separated configurations as the minimum allowable spacing increases.

For each set of parameters, ten random realizations of the RVE were generated, and the average effective stiffness was computed. The results for clustered and equilibrium fiber distributions, as well as for idealized hexagonal packing, are presented in Fig. 10. Each RVE contained 250 fibers at a volume fraction of 0.65. The constituent elastic properties used in the simulations corresponded to a fiber-to-matrix stiffness ratio of 26.5, with Poisson's ratios of 0.35 for the matrix and 0.25 for the fibers. Fig. 10 reports the mean effective stiffness values, normalized by the Young's modulus of the matrix, together with the standard deviation for each set of RVEs.

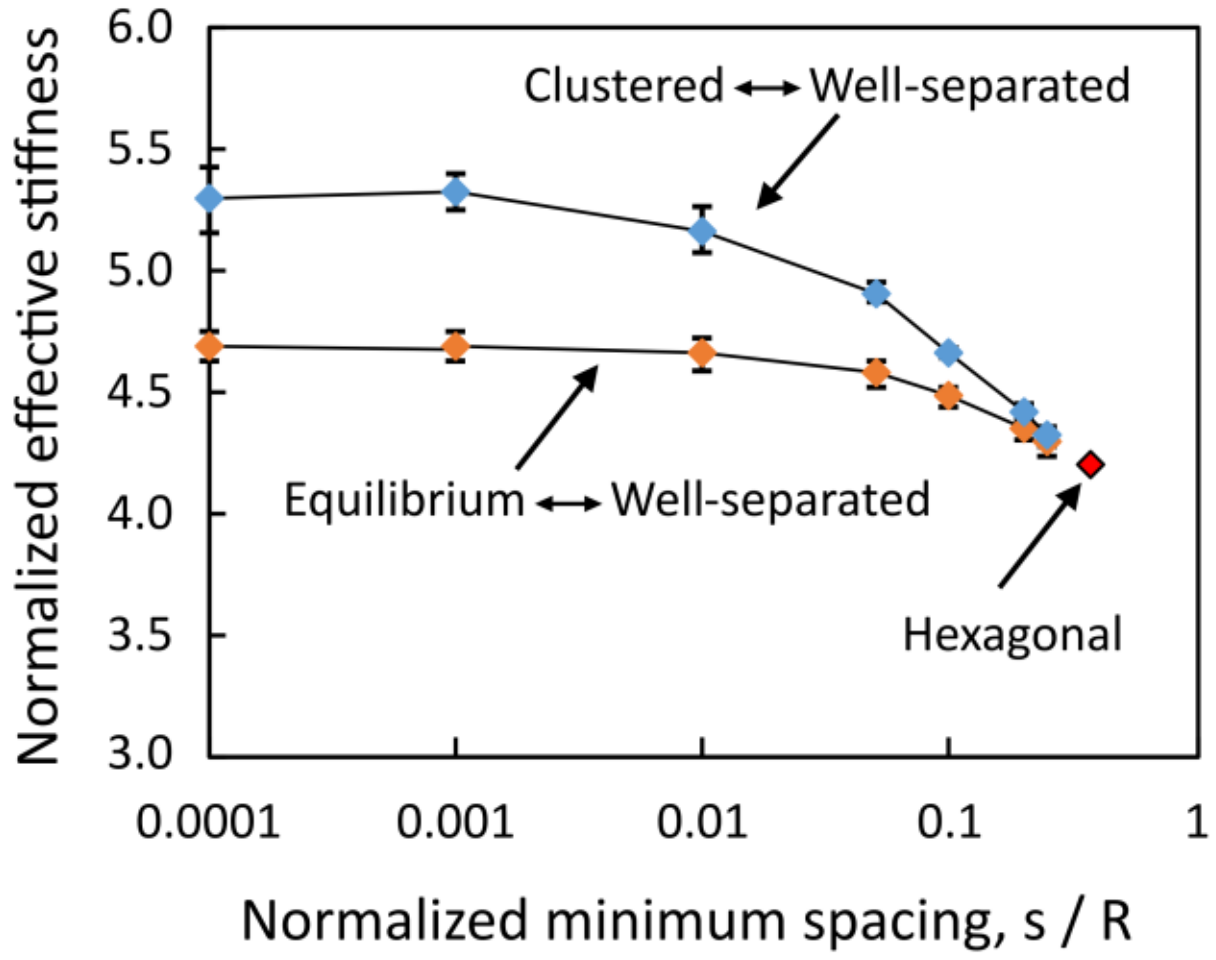

Fig. 10. Effect of the minimum interparticle spacing on the effective stiffness for clustered (blue diamonds) and equilibrium (orange diamonds) distributions of fibers (the stiffness for the hexagonal arrangement of fibers is shown as a red diamond)

The results indicate that clustered fiber distributions exhibit substantially higher effective stiffness compared to equilibrium-like distributions. As the minimum allowable interparticle distance decreases, the effective stiffness approaches a plateau; for values below approximately $0.01R$, the calculated stiffness becomes effectively insensitive to further reductions in the minimum spacing. Conversely, increasing the admissible distance between fibers leads to a gradual reduction in stiffness for both clustered and equilibrium distributions, ultimately converging toward the minimum value associated with the hexagonal packing configuration. A similar trend has been reported for the shear modulus of composites reinforced with spherical particles [35].

Based on these observations, a minimum allowable distance of $0.01R$ was adopted for subsequent simulations. At this value, further reduction does not significantly improve accuracy, while the computational cost stays within practical limits.

### 6.3. Statistical upper and lower envelopes for random composites

Finite element simulations demonstrate that, for a fixed fiber volume fraction, clustered fiber distributions exhibit the highest effective stiffness, whereas equilibrium-type distributions yield the lowest values. By generating a large number of microstructures using the SRM algorithm with different swelling and migration parameters, a cloud of effective stiffness values can be obtained, as shown in Fig. 11. Each point in this cloud corresponds to a particular fiber arrangement and represents one possible stiffness value of a random fiber-reinforced composite. All simulations in Fig. 11 were performed for a fiber-to-matrix stiffness ratio of 25 and Poisson's ratios of 0.25 and 0.35 for fibers and matrix, respectively.

The upper part of the cloud is formed by clustered configurations at various fiber volume fractions, while the lower part corresponds to equilibrium distributions, and they form statistical upper and lower envelopes for random composites. Intermediate points arise from partially clustered or partially uniform arrangements.

The stiffness of the hexagonal arrangement of monodisperse fibers, shown as a red dashed line in Fig. 11, provides a good estimate of the lowest transverse stiffness among regular and statistically uniform monodisperse microstructures. Since equilibrium configurations lie just above the stiffness of the hexagonal packing, most random composites generated from near-Poisson initial conditions are expected to fall within this envelope, with the exception of highly non-uniform, strongly clustered microstructures.

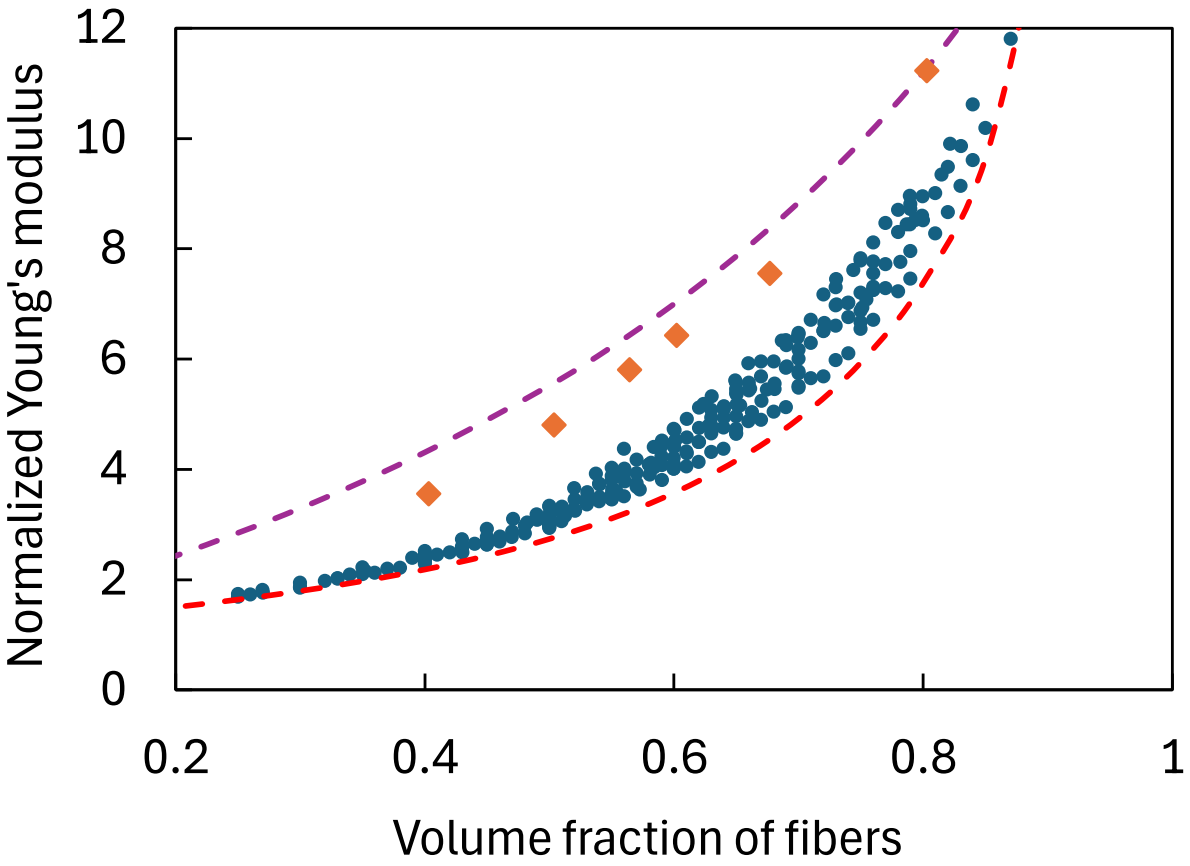

Fig. 11. Effective stiffness of SRM generated random microstructures (points) bounded from below by the stiffness of the hexagonal packing (red dashed line) and from above by the two step Mori–Tanaka estimate (violet dashed line). The stiffness of the constructed isotropic "hexagonal with holes" configurations is shown as diamonds.

In these simulations all microstructures were generated from initial seed positions close to a Poisson point process, which limits the maximum degree of clusterization achievable within the SRM framework. More strongly clustered configurations, and therefore higher stiffness values, can be obtained for highly non-uniform distributions of fibers, as will be demonstrated below.

Our numerical results indicate that the maximum effective stiffness is achieved by developing a continuous, percolating network of closely spaced fibers, within which the matrix appears as isolated soft islands. To approximate the upper range of stiffness attainable by clustered fiber-reinforced microstructures, a two-step homogenization scheme was employed. First, the stiff phase was modeled as a homogenized composite with the effective properties of a hexagonal fiber arrangement at a fiber volume fraction of 0.9. Second, soft inclusions with the properties of the matrix were embedded into this effective medium using the Mori–Tanaka scheme. This construction represents a random composite in which the stiff phase forms a continuous load-bearing skeleton, while the soft phase appears as isolated inclusions. Although this model does not constitute a strict theoretical upper bound, it provides a realistic estimate of the stiffest random microstructures compatible with the given phase properties and volume fractions, and is shown in Fig. 11 by violet dashed line.

Regular stiff isotropic microstructures can be constructed by starting from a hexagonal close packing and removing fibers in a symmetric pattern. The effective stiffness of these "depleted hexagonal" structures was found to be significantly higher than that of statistically random microstructures generated by the SRM algorithm, but still lower than the value predicted by the two-step Mori–Tanaka scheme. Examples of constructed stiff regular lattice structures are shown in Fig. 12, and corresponding effective stiffness is presented in Fig. 11 as orange diamonds.

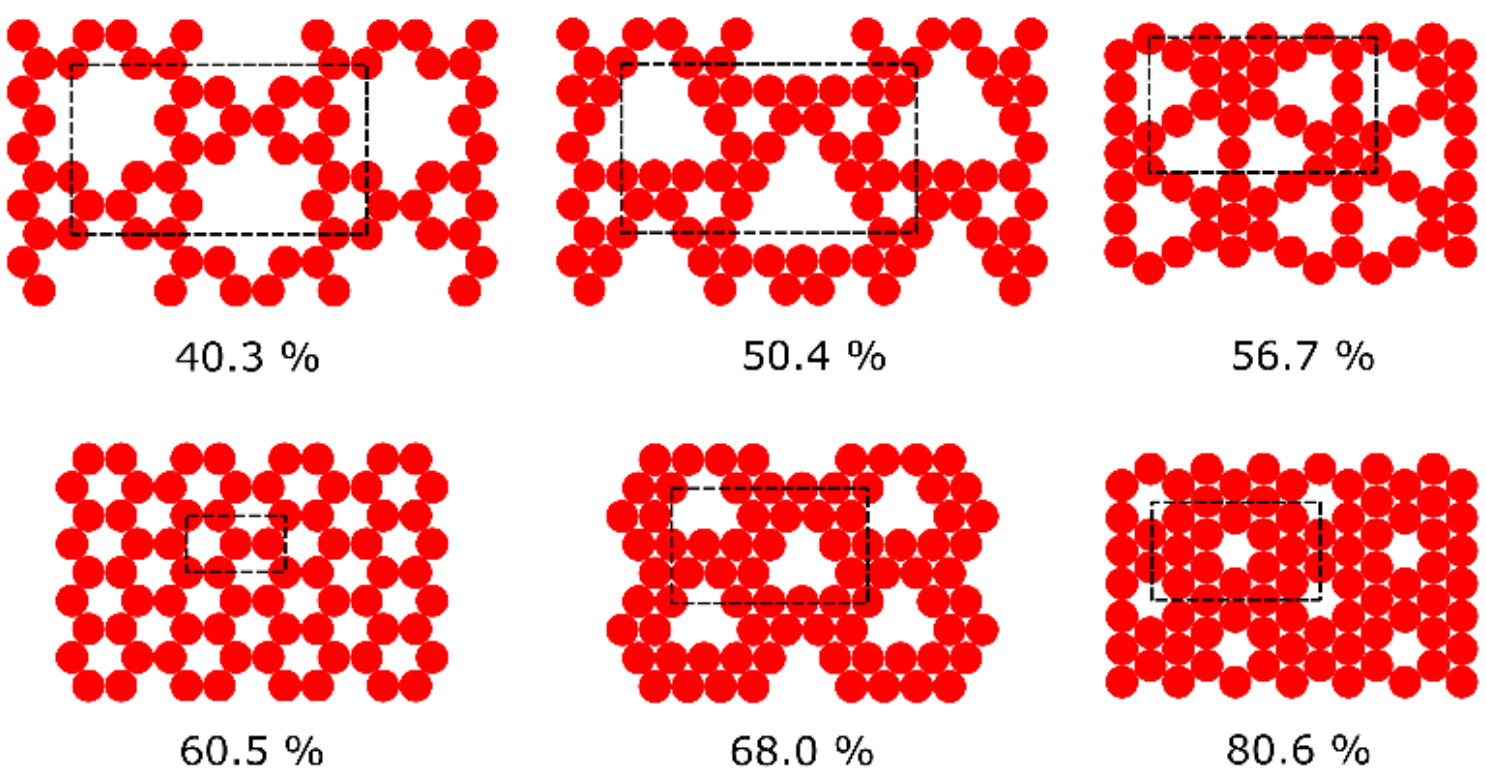

Fig. 12. Regular isotropic microstructures with different volume fractions generated by deleting fibers from a hexagonal close-packed arrangement in symmetric patterns. Periodic unit cells are shown by dashed rectangles.

## 6.4. Effect of the mean nearest neighbor distance

The transition from clustered to equilibrium fiber distributions can be characterized by a single quantitative descriptor: the mean nearest neighbor (MNN) distance. Fig. 13 presents the dependence of the effective stiffness on the normalized MNN distance $d/R$ for fiber volume fractions ranging from 40% to 79%. For each volume fraction, several RVEs were analyzed with varying degrees of spatial randomness, from highly clustered configurations, corresponding to the smallest MNN distances, to equilibrium-like distributions, which exhibit the largest MNN distances.

The results demonstrate a clear linear relationship between the effective stiffness and the MNN distance. Clustered microstructures, characterized by small nearest neighbor distances, exhibit the highest stiffness, whereas more random (equilibrium-like) distributions show progressively lower stiffness, reaching a minimum at the equilibrium state. Similar trend for the shear modulus of particle-reinforced composites has been reported in [36]. The stiffness values shown in Fig. 13 correspond to those presented in Fig. 11 for selected fiber volume fractions, and the dotted lines represent linear fits for each group of data points.

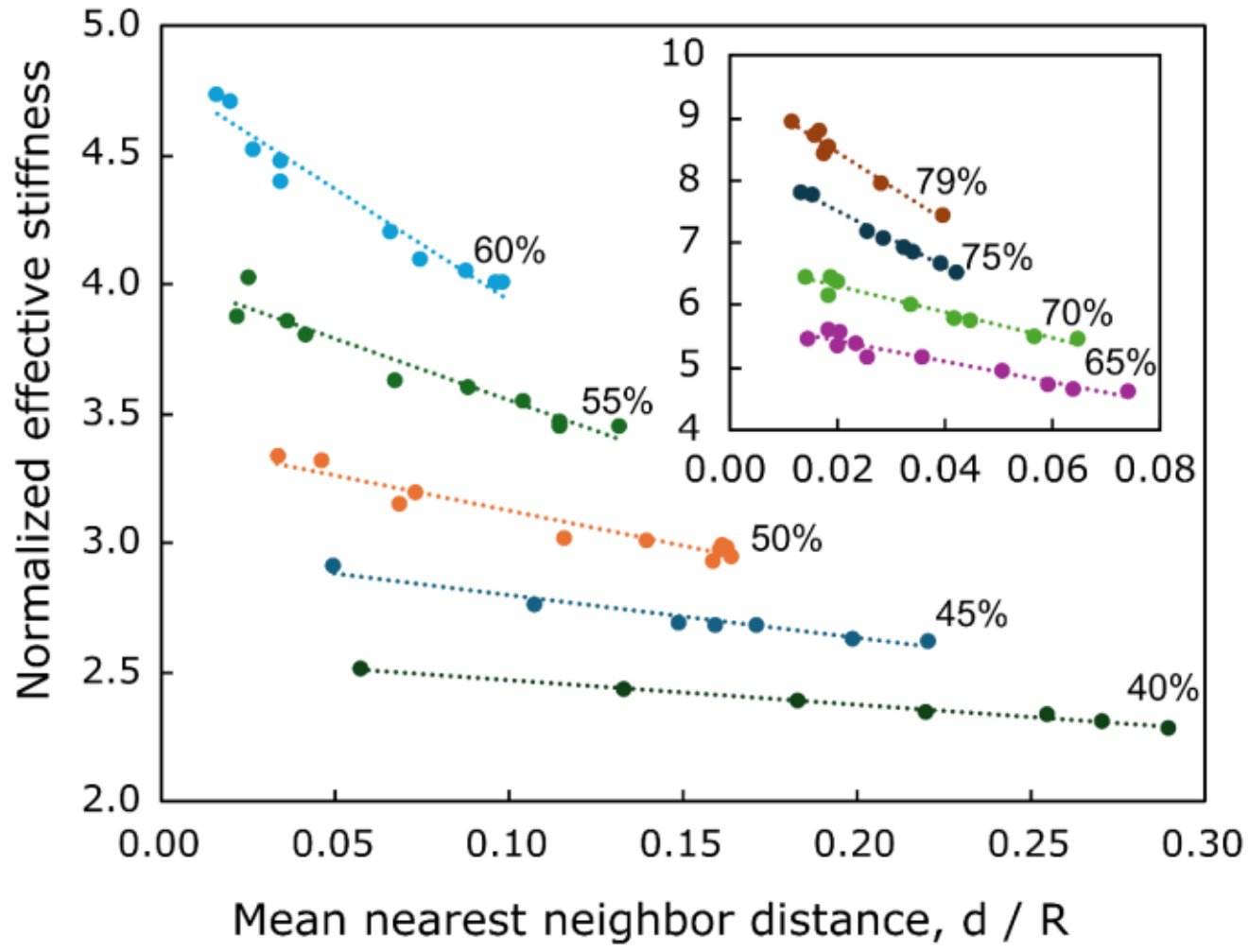

Fig. 13. Effect of the mean nearest neighbor distance between fibers on the effective stiffness for various volume fractions of fibers.

It should be noted, however, that the results presented above were obtained for a specific class of fiber arrangements generated from a near-Poisson distribution of seed points using the SRM procedure. These "random-clustered" configurations exhibit only a limited degree of clusterization, constrained by the underlying stochastic initialization and the random movements during the swelling phase. A higher level of clusterization can be achieved either by starting from a deliberately non-uniform initial distribution of fibers or by imposing spatial restrictions on the random migration step within the SRM algorithm.

The latter approach was adopted in this work to generate more clustered microstructures. Restricted migration regions in the form of circular domains or elongated thick "needle-shaped" regions were prescribed, producing two distinct types of clustered configurations, illustrated in Fig. 14 for an average fiber volume fraction of 65%. After generating clustered arrangements with prescribed volume fiber fraction, a slow relaxation process was applied to adjust the microstructures so that all three configurations shared the same mean nearest neighbor distance as the "random-clustered" arrangement of fibers shown in Fig. 14a. This allowed a direct comparison of their effective elastic properties.

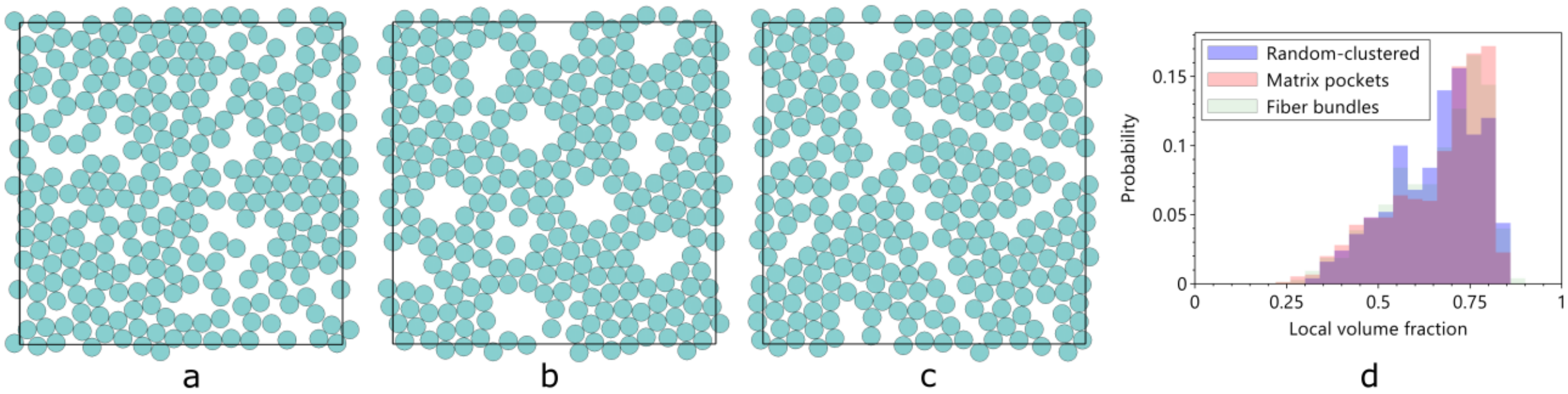

Fig. 14. Examples of random-clustered fiber arrangements generated using the standard SRM procedure (a) and using restricted migration regions, producing "matrix pockets" (b) and "fiber bundles" (c) configurations. Panel (d) shows the corresponding local volume fraction distributions for these microstructures.

Geometrically, these configurations resemble either matrix pockets embedded within a dense, continuous network of closely packed fibers (Fig. 14b) or tightly packed fiber bundles separated by narrow matrix channels (Fig. 14c). Both differ markedly from the random-clustered structures produced by the standard SRM process, however having the same MNN distances as random-clustered distribution, and showing quite similar local volume fraction distributions (Fig. 14d). Correspondingly, their effective stiffnesses bracket the stiffness of the random-clustered configuration: the "matrix pockets" morphology yields a higher stiffness, while the "fiber bundles" morphology yields a lower stiffness.

These examples clearly demonstrate that such strongly clustered microstructures cannot be characterized by a single scalar descriptor such as the mean nearest neighbor distance. More sophisticated measures that capture the topology and connectivity of the stiff phase are required. Nevertheless, the effective stiffnesses of the two extreme clustered configurations in Fig. 14b-c differ by less than 10%, and the stiffness of the random-clustered configuration (Fig. 14a) lies approximately midway between them. This indicates that, for sufficiently random or moderately clustered fiber arrangements, the mean nearest neighbor distance serves as an efficient and practically useful single numerical descriptor for estimating the effective stiffness of unidirectional composites.

## 6.5. Effect of the distribution of fibers on the transverse tensile strength

To investigate the influence of fiber distribution on the transverse tensile strength of a unidirectional composite, an initial clustered configuration containing 200 fibers with diameters of 24 μm and a fiber volume fraction of 0.6 was generated (Fig. 15a). Starting from this configuration, relaxation iterations were performed using a zero swelling rate and small migration rate. This procedure gradually reduced the degree of clustering and produced increasingly random arrangements (Fig. 15b-c), ultimately converging to an equilibrium-like distribution (Fig. 15d). In this manner, a sequence of RVEs was obtained that represents a continuous transition from a strongly clustered microstructure to an equilibrium distribution through several intermediate states.

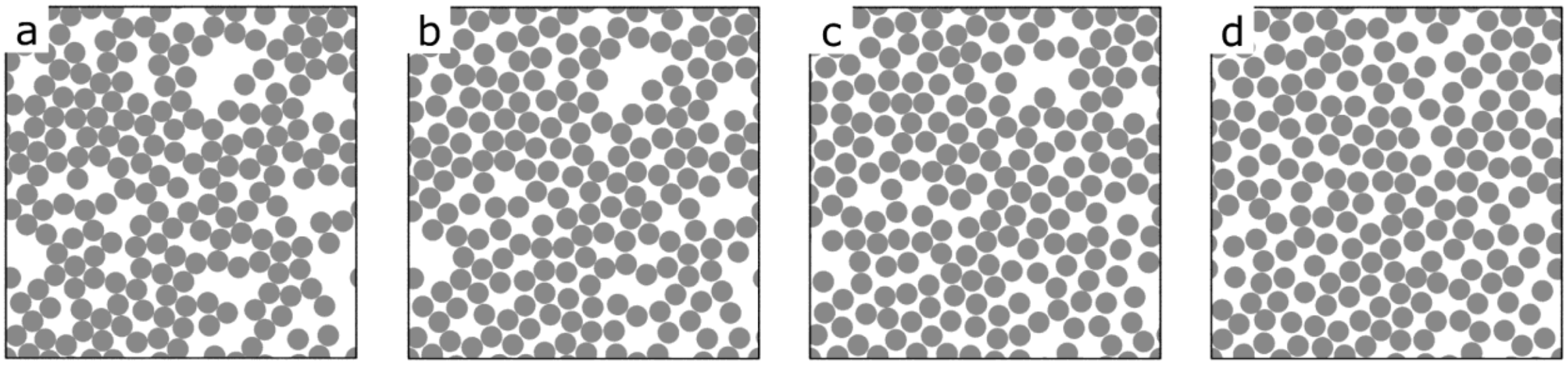

Fig. 15. Transition from clustered (a) to equilibrium (d) distribution of fibers through several intermediate states (b, c).

The transverse tensile strength of the composite was estimated following the ductile fracture modeling approach described in [37]. Finite element simulations of transverse tension were performed in ABAQUS using a two-dimensional plane strain formulation. The matrix and fibers were discretized with reduced integration quadrilateral elements (CPE4R), while interfacial debonding was represented using cohesive elements (COH2D4). The cohesive interface was modeled with a linear elastic traction-separation law followed by linear damage evolution. Damage initiation occurred when the quadratic stress criterion reached its critical value. The fibers were treated as linear elastic with typical glass-fiber properties, whereas the matrix was modeled using the built-in linear Drucker–Prager plasticity model combined with a ductile fracture criterion. An explicit dynamic solver was employed to avoid convergence difficulties, and mass scaling was applied to reduce computational time.

The effective transverse tensile properties were evaluated for a typical glass-fiber unidirectional composite. All material and interface parameters were taken from [37]. The Young's modulus of the fibers and matrix was 74 GPa and 3.35 GPa, respectively, with corresponding Poisson's ratios of 0.20 and 0.35. The matrix exhibited a tensile strength of 80 MPa. The fiber-matrix interface strength was 53 MPa in tension and 75 MPa in shear, with fracture toughness values of 2 J/m² and 5 J/m², respectively.

The progressive failure simulations show that damage typically initiates within the narrow matrix ligaments separating adjacent fibers, where pronounced stress concentrations develop. Once local debonding begins, damage propagates through interfacial debonding between the fibers and the matrix. As these debonded regions grow and coalesce, a macroscopic crack forms, ultimately leading to global failure of the RVE. Because clustered configurations contain more closely spaced fibers, they exhibit higher local stress peaks and therefore lower applied stresses at the onset of damage, a trend clearly reflected in the corresponding stress-strain curves. In contrast, near-equilibrium configurations, characterized by larger inter-fiber distances, show higher initial damage stresses as well as higher ultimate tensile strength. Fig. 16 illustrates typical crack paths for clustered (Fig. 16a) and equilibrium (Fig. 16b) fiber arrangements near complete failure, with representative stress-strain responses shown in Fig. 16c. Similar observations were reported in [7].

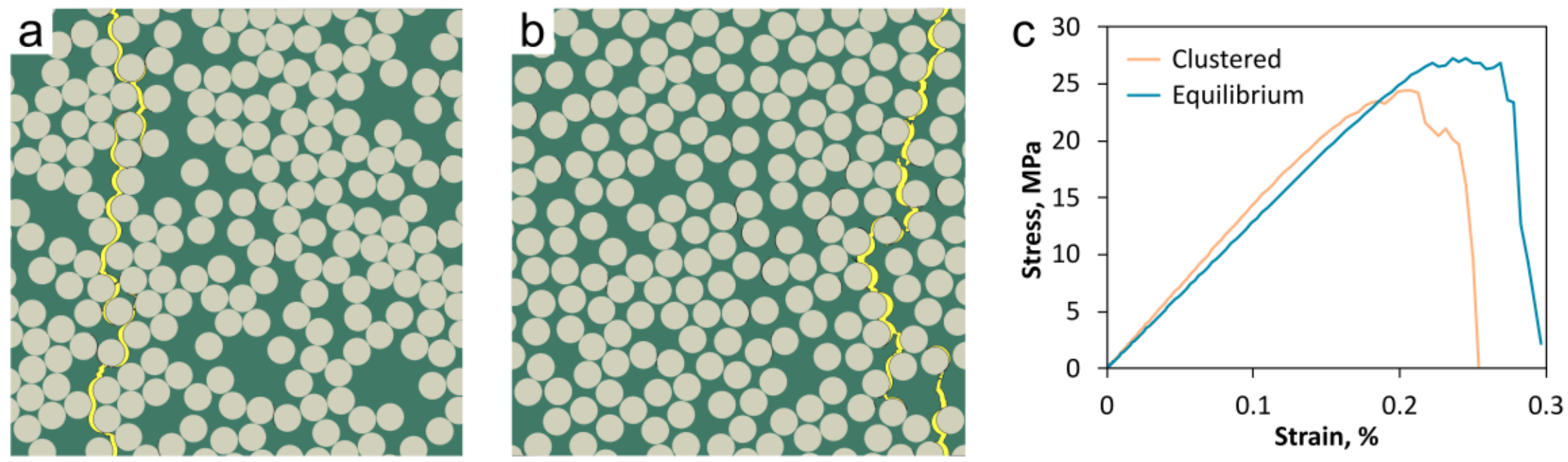

Fig. 16. Typical failure modes of clustered (a) and equilibrium (b) arrangements of fibers and corresponding numerical stress-strain curves (c).

Three series of RVEs representing a gradual transition from clustered to equilibrium fiber distributions were analyzed. For each RVE, the effective transverse stiffness and tensile strength were computed in two orthogonal directions, and the corresponding average values were used for comparison. The results are summarized in Fig. 17, where the averaged data points (solid markers) and fitted linear trendlines (dotted lines) illustrate the dependence of both stiffness and strength on the mean nearest neighbor distance. All values were normalized by the matrix Young's modulus and matrix tensile strength, respectively.

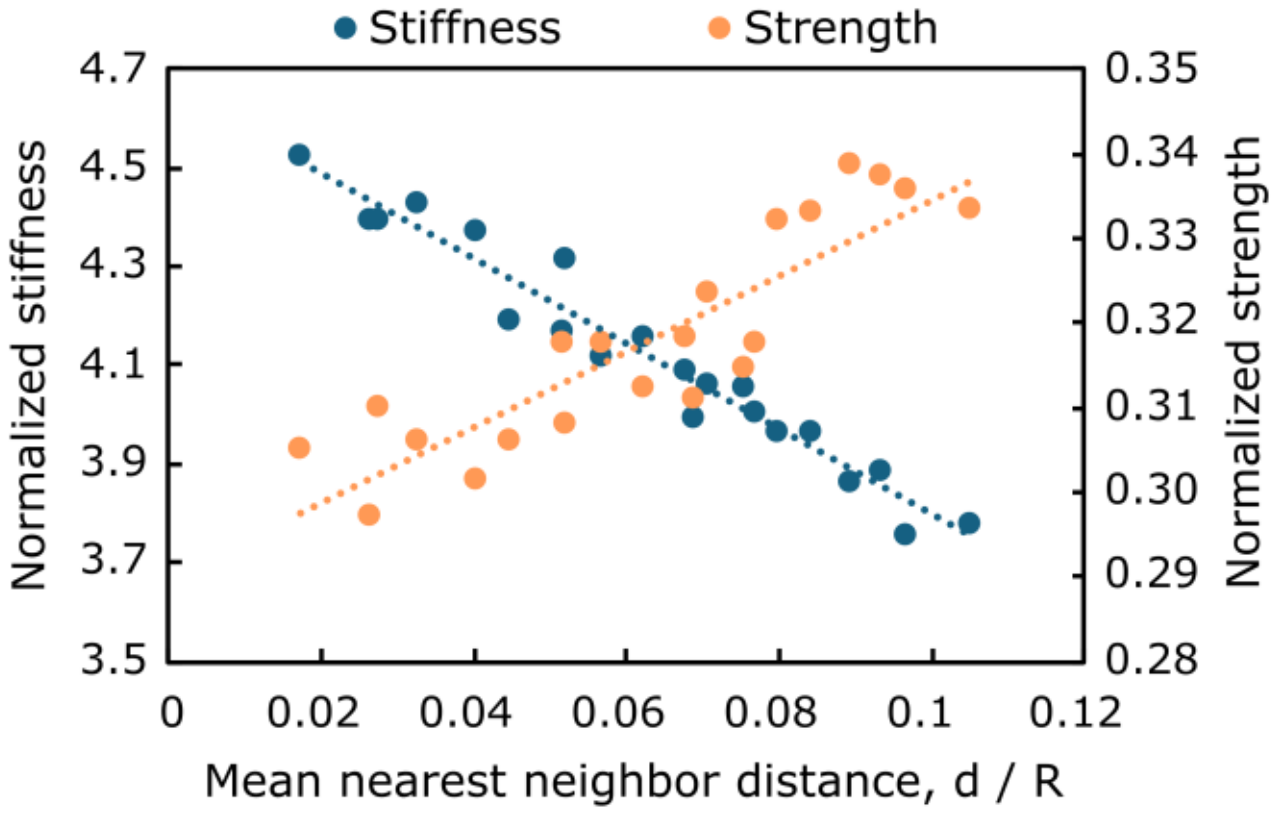

Fig. 17. Dependence of the effective transverse stiffness and strength of a unidirectional composite on the mean nearest neighbor's distance of fibers at a fiber volume fraction of 60%

The obtained results indicate that, similar to the effective stiffness, the simulated transverse tensile strength of the unidirectional composite varies linearly with the mean nearest neighbor distance. However, in contrast to the stiffness, the transverse strength increases as the mean nearest neighbor distance increases. For a typical glass-fiber/epoxy composite with a fiber volume fraction of 60%, the difference in effective transverse properties between clustered and equilibrium-like distributions can reach approximately 15%.

Although the transverse tensile strength was evaluated using a detailed progressive failure model, these results should be interpreted with greater caution than the stiffness predictions. Strength calculations inherently involve more assumptions, particularly regarding matrix plasticity, interface behavior, and damage evolution, and consequently exhibit larger scatter across nominally similar RVEs. As a result, the conclusions drawn for transverse strength are more speculative than those for stiffness and should be viewed as indicative trends rather than precise quantitative predictions.

# 7. Conclusions

A Swelling and Random Migration algorithm was used to generate realistic fiber distributions that replicate the spatial statistics observed in real composites. Several statistical descriptors were used to characterize the distribution of fibers, including nearest neighbor distribution, Voronoi neighbors distribution, local volume fraction, Ripley's $K$-function, and pair distribution function.

Polished micro-images of composite samples were used to compare the generated distributions of fibers with experimentally measured distributions. It was found that with a proper choice of parameters, the SRM algorithm is able to generate random distributions of fibers that are statistically equivalent to real distributions, i.e., all calculated statistical descriptors of generated RVE closely resemble the descriptors of real composite samples. The calculated effective stiffness of UD composite having statistically equivalent distributions of fibers practically coincides with that of experimental distributions reconstructed from micro-images of studied samples.

Variations in the minimum allowable interparticle distance were found to strongly affect the effective stiffness of unidirectional composites: increasing the minimum spacing between fibers leads to progressively lower stiffness values, approaching the limit defined by hexagonal packing. For sufficiently small spacings (below about 0.01 of fiber's radius), the effective stiffness becomes insensitive to further reductions, indicating a practical lower bound for numerical modeling.

Finite element simulations reveal that, for a fixed fiber volume fraction, clustered microstructures consistently yield the highest effective stiffness, whereas equilibrium-type arrangements produce the lowest values. By generating a large number of microstructures with varying volume fractions and degrees of clusterization, statistical upper and lower envelopes for random composites are obtained. The upper envelope corresponds to clustered configurations, while the lower envelope is defined by equilibrium distributions.

A single scalar descriptor, the mean nearest neighbor distance, is introduced to quantify the transition from clustered to equilibrium fiber distributions. For a specific class of microstructures, namely sufficiently random fiber arrangements, both stiffness and transverse tensile strength exhibit linear dependence on the MNN distance, with opposite trends: effective stiffness decreases, whereas transverse strength increases as MNN distance increases. This shows that the MNN distance acts as an efficient and practically useful scalar descriptor for estimating the effective transverse properties of unidirectional composites.

**Acknowledgement**

The studies were performed within the project "Graded Interphases for Enhanced Dielectric and Mechanical Strength of Fiber Reinforced Composites "[GRADIENT] project carried out under the M ERA.NET 3 scheme (European Union's Horizon 2020 research and innovation program under grant agreement No 958174) and co-financed with tax funds on the basis of the budget passed by the Saxon state parliament), grant No. 360049, and grant No. ES RTD/2022/16 from the Latvian Council of Science.

**CRediT authorship contribution statement**

**Sergejs Tarasovs**: Writing - Original Draft, Writing - Review & Editing, Conceptualization, Methodology, Software, Validation, Formal analysis, Investigation, Data Curation. **Janis Modniks**: Writing - Original Draft, Writing - Review & Editing, Software, Validation, Investigation. **Andrea Bercini Martins**: Writing - Review & Editing, Methodology, Investigation. **Christina Scheffler**: Writing - Review & Editing, Project administration, Funding acquisition. **Janis Andersons**: Writing - Original Draft, Writing - Review & Editing, Conceptualization, Formal analysis, Resources, Supervision, Project administration, Funding acquisition.

**Declaration of competing interest**

The authors declare that they have no known competing financial interests or personal relationships that could have appeared to influence the work reported in this paper.

**Data availability**

Data will be made available on request.

# References

[1] R. Pyrz, Quantitative description of the microstructure of composites. Part I: Morphology of unidirectional composite systems, Compos. Sci. Technol. 50(2) (1994) 197-208. https://doi.org/10.1016/0266-3538(94)90141-4
[2] T.A. Dutra, R.T.L. Ferreira, H.B. Resende, L.M. Oliveira, B.J. Blinzler, L.E. Asp, Identification of Representative Equivalent Volumes on the Microstructure of 3D-Printed Fiber-Reinforced Thermoplastics Based on Statistical Characterization, Polymers 14(5) (2022) 972. https://doi.org/10.3390/polym14050972
[3] A.A. Gusev, P.J. Hine, I.M. Ward, Fiber packing and elastic properties of a transversely random unidirectional glass/epoxy composite, Compos. Sci. Technol. 60(4) (2000) 535-541. https://doi.org/10.1016/S0266-3538(99)00152-9
[4] T.J. Vaughan, C.T. McCarthy, A combined experimental–numerical approach for generating statistically equivalent fibre distributions for high strength laminated composite materials, Compos. Sci. Technol. 70(2) (2010) 291-297. https://doi.org/10.1016/j.compscitech.2009.10.020
[5] B.A. Bednarcyk, J. Aboudi, S.M. Arnold, Analysis of fiber clustering in composite materials using high-fidelity multiscale micromechanics, Int. J. Solids Struct. 69-70 (2015) 311-327. https://doi.org/10.1016/j.ijsolstr.2015.05.019
[6] M.J. Schey, T. Beke, L. Appel, S. Zabler, S. Shah, J. Hu, F. Liu, M. Maiaru, S. Stapleton, Identification and Quantification of 3D Fiber Clusters in Fiber-Reinforced Composite Materials, JOM 73(7) (2021) 2129-2142. https://doi.org/10.1007/s11837-021-04703-0
[7] X. Pang, F. Huang, F. Zhu, S. Zhang, Y. Wang, X. Chen, Progressive failure characteristics of unidirectional FRP with fiber clustering, Compos. Struct. 280 (2022) 114880. https://doi.org/10.1016/j.compstruct.2021.114880

[8] J.F. Husseini, E.J. Pineda, S.E. Stapleton, Generation of artificial 2-D fiber reinforced composite microstructures with statistically equivalent features, Compos. Part A Appl. Sci. Manuf. 164 (2023) 107260. https://doi.org/10.1016/j.compositesa.2022.107260
[9] H.N. Petersen, Y. Kusano, P. Brøndsted, K. Almdal, Preliminary characterization of glass fiber sizing, Proceedings of the Risø international symposium on materials science, Risø National Laboratory, 2013, pp. 333-340.
[10] A.E. Krauklis, A.I. Gagani, A.T. Echtermeyer, Long-Term Hydrolytic Degradation of the Sizing-Rich Composite Interphase, Coatings 9(4) (2019) 263. https://doi.org/10.3390/coatings9040263
[11] V. Cech, Plasma-polymerized organosilicones as engineered interlayers in glass fiber/polyester composites, Compos. Interfaces 14(4) (2007) 321-334. https://doi.org/10.1163/156855407780452850
[12] V. Romanov, S.V. Lomov, Y. Swolfs, S. Orlova, L. Gorbatikh, I. Verpoest, Statistical analysis of real and simulated fibre arrangements in unidirectional composites, Compos. Sci. Technol. 87 (2013) 126-134. https://doi.org/10.1016/j.compscitech.2013.07.030
[13] K.C. Liu, A. Ghoshal, Validity of random microstructures simulation in fiber-reinforced composite materials, Compos. Part B Eng. 57 (2014) 56-70. https://doi.org/10.1016/j.compositesb.2013.08.006
[14] W. Ge, L. Wang, Y. Sun, X. Liu, An efficient method to generate random distribution of fibers in continuous fiber reinforced composites, Polym. Compos. 40(12) (2019) 4763-4770. https://doi.org/10.1002/pc.25344
[15] S. Ghosh, Z. Nowak, K. Lee, Quantitative characterization and modeling of composite microstructures by Voronoi cells, Acta Mater. 45(6) (1997) 2215-2234. https://doi.org/10.1016/S1359-6454(96)00365-5
[16] S.A. Elnekhaily, R. Talreja, Damage initiation in unidirectional fiber composites with different degrees of nonuniform fiber distribution, Compos. Sci. Technol. 155 (2018) 22-32. https://doi.org/10.1016/j.compscitech.2017.11.017
[17] M.H. Nagaraj, M. Schey, T. Beke, S.E. Stapleton, M. Maiaru, Influence of microstructural variabilities on the mechanical properties of fiber-reinforced composites accounting for manufacturing effects, Compos. Struct. 373 (2025) 119631. https://doi.org/10.1016/j.compstruct.2025.119631
[18] M.V. Pathan, V.L. Tagarielli, S. Patsias, P.M. Baiz-Villafranca, A new algorithm to generate representative volume elements of composites with cylindrical or spherical fillers, Compos. Part B Eng. 110 (2017) 267-278. https://doi.org/10.1016/j.compositesb.2016.10.078
[19] A.R. Melro, P.P. Camanho, S.T. Pinho, Influence of geometrical parameters on the elastic response of unidirectional composite materials, Compos. Struct. 94(11) (2012) 3223-3231. https://doi.org/10.1016/j.compstruct.2012.05.004
[20] T. Zhang, Y. Yan, A comparison between random model and periodic model for fiber-reinforced composites based on a new method for generating fiber distributions, Polym. Compos. 38(1) (2017) 77-86. https://doi.org/10.1002/pc.23562
[21] S. Tarasovs, Efficient generation of large-scale non-equilibrium distributions of particles, arXiv (arXiv:2605.18254) (2026). https://doi.org/10.48550/arXiv.2605.18254
[22] J.L.P. Vila-Chã, B.P. Ferreira, F.M.A. Pires, An adaptive multi-temperature isokinetic method for the RVE generation of particle reinforced heterogeneous materials, Part I: Theoretical formulation and computational framework, Mech. Mater. 163 (2021) 104069. https://doi.org/10.1016/j.mechmat.2021.104069
[23] M.D. Rintoul, S. Torquato, Reconstruction of the Structure of Dispersions, J. Colloid Interface Sci. 186(2) (1997) 467-476. https://doi.org/10.1006/jcis.1996.4675
[24] C. Geuzaine, J.-F. Remacle, Gmsh: A 3-D finite element mesh generator with built-in pre- and post-processing facilities, Int. J. Numer. Methods Eng. 79(11) (2009) 1309-1331. https://doi.org/10.1002/nme.2579

[25] A.R. Melro, P.P. Camanho, S.T. Pinho, Generation of random distribution of fibres in long-fibre reinforced composites, Compos. Sci. Technol. 68(9) (2008) 2092-2102. https://doi.org/10.1016/j.compscitech.2008.03.013
[26] H. Ghayoor, S.V. Hoa, C.C. Marsden, A micromechanical study of stress concentrations in composites, Compos. Part B Eng. 132 (2018) 115-124. https://doi.org/10.1016/j.compositesb.2017.09.009
[27] D. Trias, J. Costa, A. Turon, J.E. Hurtado, Determination of the critical size of a statistical representative volume element (SRVE) for carbon reinforced polymers, Acta Mater. 54(13) (2006) 3471-3484. https://doi.org/10.1016/j.actamat.2006.03.042
[28] J. Zeman, M. Šejnoha, Numerical evaluation of effective elastic properties of graphite fiber tow impregnated by polymer matrix, J. Mech. Phys. Solids 49(1) (2001) 69-90. https://doi.org/10.1016/S0022-5096(00)00027-2
[29] R.K. Everett, J.H. Chu, Modeling of Non-Uniform Composite Microstructures, J. Compos. Mater. 27(11) (1993) 1128-1144. https://doi.org/10.1177/002199839302701105
[30] M. Suquet P, Elements of Homogenization Theory for Inelastic Solid Mechanics, in: E. Sanchez-Palencia, A. Zaoui (Eds.), Homogenization Techniques for Composite Media, Springer-Verlag, Berlin, Germany, 1987, pp. 193-287.
[31] S. Li, On the nature of periodic traction boundary conditions in micromechanical FE analyses of unit cells, IMA J. Appl. Math. 77(4) (2012) 441-450. https://doi.org/10.1093/imamat/hxr024
[32] Z. Xia, C. Zhou, Q. Yong, X. Wang, On selection of repeated unit cell model and application of unified periodic boundary conditions in micro-mechanical analysis of composites, Int. J. Solids Struct. 43(2) (2006) 266-278. https://doi.org/10.1016/j.ijsolstr.2005.03.055
[33] M. Jiang, K. Alzebdeh, I. Jasiuk, M. Ostoja-Starzewski, Scale and boundary conditions effects in elastic properties of random composites, Acta Mech. 148(1) (2001) 63-78. https://doi.org/10.1007/BF01183669
[34] S. Hazanov, C. Huet, Order relationships for boundary conditions effect in heterogeneous bodies smaller than the representative volume, J. Mech. Phys. Solids 42(12) (1994) 1995-2011. https://doi.org/10.1016/0022-5096(94)90022-1
[35] A.A. Gusev, Controlled accuracy finite element estimates for the effective stiffness of composites with spherical inclusions, Int. J. Solids Struct. 80 (2016) 227-236. https://doi.org/10.1016/j.ijsolstr.2015.11.006
[36] P.-Y. Mechin, A. Borras, K. Cottard, V. Keryvin, A unified method to generate representative volume elements with tailored random fibre arrangements to estimate the shear and transverse behaviours of unidirectional continuous fibres composite plies, J. Compos. Mater. (2024) 00219983241300144. https://doi.org/10.1177/00219983241300144
[37] A. Sharma, S. Daggumati, Computational micromechanical modeling of transverse tensile damage behavior in unidirectional glass fiber-reinforced plastic composite plies: Ductile versus brittle fracture mechanics approach, Int. J. Damage Mech. 29(6) (2019) 943-964. https://doi.org/10.1177/1056789519894379